\RequirePackage{fix-cm}
\documentclass[smallextended]{svjour3}       % onecolumn (second format)
\smartqed  % flush right qed marks, e.g. at end of proof
\usepackage{graphicx}
\usepackage{amsmath}
\DeclareMathOperator{\arsinh}{arsinh}
\DeclareMathOperator{\Sk}{ S_{\kappa}}
\DeclareMathOperator{\ku}{ u_{\kappa} }
\DeclareMathOperator{\kexp}{ \exp_{ \kappa } }
\DeclareMathOperator{\kln}{ \ln_{\kappa} }
\newcommand{\ave}[1]{\left\langle #1 \right\rangle}
\journalname{Journal of Statistical Physics}
\begin{document}

\title{On the canonical distributions of a thermal particle in the weakly confining potential of special type
\thanks{The first named author (T.W.) is supported by Japan Society for the Promotion of Science (JSPS) Grants-in-Aid for Scientific Research (KAKENHI) Grant Number JP17K05341. The third named author (H. M.) is partially supported by the JSPS Grants-in-Aid for Scientific Research (KAKENHI) Grant Number JP26108003 and JP15K04842.
%Grants or other notes
}
}


\author{Tatsuaki Wada         \and
        Antonio M. Scarfone \and
        Hiroshi Matsuzoe %etc.
}


\institute{T. Wada \at
              Region of Electrical and Electronic Systems Engineering, Ibaraki University, Nakanarusawa-cho, Hitachi 316-8511, Japan \\
              Tel.: +81-294-38-5110\\
              Fax: +81-294-38-5275\\
              \email{tatsuaki.wada.to@vc.ibaraki.ac.jp}           %  \\
           \and
           A. M. Scarfone \at
             Istituto dei Sistemi Complessi, Consiglio Nazionale delle Ricerche (ISC-CNR), c/o Politecnico di Torino, 10129 Torino, Italy
          \and
          H. Matsuzoe \at
Department of Computer Science and Engineering, Nagoya Institute of Technology, Gokiso-cho, Showa-ku, Nagoya, 466-8555, Japan
}

\date{Received: date / Accepted: date}

\maketitle

%Insert your abstract here. 
\begin{abstract}
We consider a thermal particle which is diffusing in velocity-space and in a weakly confining potential characterized by
the inverse hyperbolic sine function of the particle velocity $v$ and the control parameter $v_c$. The stationary state of the Fokker-Planck equation is shown to be a canonical probability distribution. 
Furthermore an appropriate re-parametrization relates this stationary state with the $\kappa$-deformed Gaussian.

%Include keywords, PACS and mathematical
%subject classification numbers as needed.
\end{abstract}

\section{Introduction}
\label{intro}

When one describes basic equations for physical systems, some linear constitutive relations are often used. For example, Hooke's law $F_e \propto x$ for the elastic force $F_e$ of a spring against the change in its length $x$, Stokes' law $F_S \propto v_f$ for the frictional force $F_S$ exerted on an object in a fluid with the flow velocity $v_f$, and so on. 
However, as a real spring deviates from Hooke's law, we know that   
 any linear constitutive relation describes an idealized situation, and it is merely approximated, or linearized one for describing real physical properties. In other words, non-linearity plays a crucial role to describe more realistic physical systems.

In the standard linear Fokker-Plank (FP) equation, which is one of the most fundamental equations in statistical physics, the dissipative force $F_d(v)$ is assumed to obey a linear constitutive relation $F_d(v) \propto v$. In order to handle the effect of such a velocity-proportional force, Rayleigh \cite{Rayleigh} introduced a velocity dependent potential $v^2 / 2$, which is now called Rayleigh  dissipation function. This velocity-proportional force  is considered as the velocity gradient force of the parabolic potential $v^2 / 2$ in velocity $v$-space.
It is well known that for a thermal particle diffusing in such a parabolic potential in $v$-space, the steady state solution of the corresponding linear FP equation is a Gaussian probability density function (pdf)  $p(v) \propto \exp(-\beta v^2/2)$. 
Since  any particle with high thermal energy can be captured in a region in $v$-space around the minimum of this parabolic potential $v^2 / 2$, we call it a \textit{strongly confining} potential in $v$-space. Instead of such a strongly confining potential, in this contribution we focus on  a thermally diffusing particle in a \textit{ weakly confining} potential in $v$-space defined by
\begin{align}
  U(v; v_c) := \frac{v_c^2}{2} \arsinh \left( \frac{v^2}{v_c^2} \right),
  \label{weakU}
\end{align}
with a control parameter $v_c$. In the limit of $v_c \to \infty$, this potential reduces to a parabolic potential $v^2 / 2$. Accordingly $U(v; v_c)$ can be considered as a deformation of the parabolic potential $v^2/2$, or as a deformation of the kinetic energy of a thermal particle with unit mass $m=1$.
For a finite value of $v_c$ this potential
is weak (or shallow) in order to capture a thermal particle in a region, say $R_c$, around the minimum point (the origin $v=0$ in velocity space) of this potential $U(v; v_c)$.
Consequently,
when a particle has an enough amount of thermal energy, it cannot be captured in the region $R_c$ of this potential $U(v; v_c)$. We hence call it a weakly confining potential. The corresponding FP equation is still linear in a pdf $p$ but has a nonlinear drift force caused by this nonlinear potential $U(v; v_c)$ \cite{W10}.
Introducing such a nonlinear drift force is not a new idea. In fact Lutz \cite{LR13} showed that the anomalous transport of an atom in optical lattice was well described by the linear FP equation with another special type of the nonlinear drift force, and that its stationary state is a $q$-deformed Gaussian. 

This work is originally motivated by our recent studies on the information geometric structures
of a $\kappa$-deformed exponential family \cite{SW14,WS15,WMS15,SMW18}.  
Information geometry (IG) \cite{A-book} provides us a powerful framework for studying a family of probability distributions by identifying the space of probability distributions with a differentiable manifold endowed with a Riemannian metric and an affine connection. The exponential family of probability distributions is most familiar in IG, and it
played an important role especially in early developments of IG. 
We had studied some IG structures on the $\kappa$-deformed exponential families of probability distributions, which are non-Gaussians and with heavy-tails.
For the $\kappa$-deformed exponential families, we constructed the suitable statistical manifolds and showed some information geometric structures such as $\kappa$-generalized Fisher metrics, $\theta$- and $\eta$-potentials, dually-flat structures, $\kappa$-generalized divergence functions, and so on \cite{SW14,WS15,WMS15}.

In recent years much attention has been paid for studying the statistical physics concerning on some deformed exponential families of probability distributions. Some of them are Tsallis' $q$-deformed exponential \cite{Tsallis} and Kaniadakis' $\kappa$-deformed exponential families \cite{KS02,K02,K05} in non-extensive statistical mechanics.
The $\kappa$-deformed exponential function is defined by
\begin{align}
  \kexp(x) \equiv \left( \kappa x + \sqrt{1 + \kappa^2 x^2} \right)^{\frac{1}{\kappa}},
 \label{k-exp}
\end{align}
for a real deformed parameter $\kappa$. The $\kappa$-deformed exponential function and its inverse function, i.e., $\kappa$-deformed logarithmic function, are
important ingredients of the generalized statistical physics based on $\kappa$-entropy 
\cite{KS02,K02,K05}. 
The statistical physics based on the $\kappa$-deformed functions has been developed over a decade.
The review article \cite{K13} summarizes the theoretical foundations and mathematical formalism generated by the $\kappa$-deformed functions, and provides plentiful references including many fields of applications.
Recently, the $\kappa$-deformed Fourier series \cite{S15} and Fourier transform \cite{S17} are developed, and the open question on the composition law of $\kappa$-entropy for statistically independent systems was solved \cite{KSSW17}.
Despite such efforts for clarifying  the physical situations described by the $\kappa$-deformed functions, the physical meaning of the deformed parameter $\kappa$ is still missing.

In this contribution we consider the thermal probability distributions for the weakly confining potential $U(v; v_c)$ of Eq. \eqref{weakU} in the basic framework of statistical physics. In contrast to the well-known standard case of Gaussian distribution for strongly confining potential $v^2 /2$, it is found that the quasi-equilibrium thermal probability distribution for this weakly confining potential $U(v; v_c)$ is non-Gaussian with heavy-tails.
The corresponding FP equation of the thermal probability distribution for this weakly confining potential describes an anomalous diffusion (or anomalous transport) in a parameter region in which the second moment $\ave{v^2}$ may diverge \cite{W10}.

In addition, we relate the canonical distribution of a thermal particle in this weakly confining potential to the $\kappa$-exponential distribution ($\kappa$-deformed Gaussian),
\begin{align}
   p(v) \propto \kexp \left( -\beta \frac{v^2}{2} \right),
\end{align}
for the strongly confining potential $v^2 / 2$ by introducing a suitable re-parameterization of the control parameter $v_c$.

The paper is organized as follows. The next section provides a brief summary of the $\kappa$-deformed functions. In section III we consider a thermal particle in a confining potential and show that
for a weakly confining potential of special type \eqref{weakU}, its constitutive relation has a velocity-dependent nonlinearity. In addition, the stationary state is shown to be 
a $\kappa$-deformed Gaussian. In the region $\kappa > 2/3$ of the deformed parameter $\kappa$, our model exhibits anomalous transport. Although the mean kinetic energy $\ave{v^2/2}$ diverges, the proposed specific average does not diverge, and it satisfies a generalized equipartition relation. 
Final section is devoted our conclusion and perspective.

%%%%%%%%%%%%%%%%%%%%%%%%%%%%%%%%%%%%%%%%%%%

\section{Some $\kappa$-deformed functions}

We here briefly summarize  some $\kappa$-deformed functions and the associated useful relations
which are based on the $\kappa$-entropy $\Sk$ described by 
\begin{align}
  \Sk \equiv - \int dv \, p(v) \kln p(v) = \ave{-\kln (p)},
  \label{Sk}
\end{align}
where $v$ is the velocity of a thermal particle, and $\ave{ \cdot }$ is the expectation w.r.t $p(v)$.
Note that $\Sk$ is a $\kappa$-generalization of the Gibbs-Shannon entropy by replacing
the standard logarithm with the $\kappa$-logarithm.
Here the $\kappa$-logarithm $\kln (x)$ \cite{K02,K05} is a deformed function of 
the standard logarithm $\ln (x)$ for a real variable $x > 0$, and $\kappa$ is a real deformed parameter.
\begin{align}
 \kln(x) \equiv \frac{x^{\kappa} - x^{-\kappa}}{2 \kappa}
 = \frac{1}{\kappa} \, \sinh \big[ \kappa \ln (x) \big].
\end{align}
Its inverse function is given by Eq. \eqref{k-exp}, which is also expressed as
\begin{align}
 \exp_{\kappa} (x) 
 = \exp \left[ \frac{1}{\kappa} \arsinh \big( \kappa \, x \big) \right].
 \label{kexp}
\end{align}
In the $\kappa \to 0$ limit, the $\kappa$-exponential and the $\kappa$-logarithm reduce to the standard exponential $\exp(x)$ and
logarithm $\ln (x)$, respectively.
%\begin{align*}
% \lim_{\kappa \to 0} \kexp(x) &= \exp(x), \\
% \lim_{\kappa \to 0} \kln x &= \ln x.
%\end{align*}

We next introduce another $\kappa$-deformed function:
\begin{align}
 \ku(x) \equiv \frac{x^{\kappa} + x^{-\kappa}}{2}
 = \cosh \big[ \kappa \ln(x) \big],
\end{align}
which is the conjugate (or co-function) of $\kln x$, as similar
as that $\cos (x)$ is the co-function of $\sin (x)$. 
In the $\kappa \to 0$ limit, this $\kappa$-deformed
function reduces to the unit constant function $u_0(x)=1$.
By using this $\ku(x)$,
the derivative of the $\kappa$-exponential is expressed as
\begin{align}
  \frac{d}{dx} \kexp(x)  = \frac{ \kexp(x)}{\sqrt{1 + \kappa^2 x^2}} = \frac{ \kexp(x)}{ \ku \left[ \kexp(x) \right]},
\label{der_kexp}
\end{align}
and the derivative of $\kappa$-logarithm is expressed as
\begin{align}
  \frac{d}{dx} \kln(x)  = \frac{ \ku(x)}{x},
\end{align}
respectively.

Since all $\kappa$-deformed functions are symmetric under the sign change of the deformed parameter $\kappa$ to $-\kappa$, through out this paper we assume that the deformed parameter $\kappa$ takes a positive real value.

%%%%%%%%%%%%%%%%%%%%%%%%%%%%%%%%%%%%%%%%%%%%%%%%%%%%%%%%%%%%%%%%%%%%%%%

\section{Thermal pdf for a weakly confining potential}

We consider a thermal particle under a velocity-dependent potential $U(v)$ which acts as a confining potential in the velocity $v$-space. Don't confuse this $U(v)$ with a confining potential in the position $x$-space. Under a such velocity-dependent potential $U(v)$, the velocity of a thermal particle is limited due to the dissipative force proportional to $- d U(v) /dv$.
We call such a velocity-dependent potential $U(v)$ a \textit{confining potential} (in the velocity $v$-space). A thermal particle in this confining potential $U(v)$ is described by
the Langevin equation for the overdamped Brownian dynamics:
 \begin{align}
   \frac{d x(t)}{dt} &= v(t), \notag \\
    \frac{d v(t)}{dt} &=  -\alpha \frac{d}{d v} U(v) + \zeta(t),  
  \end{align}
where $\alpha$ denotes a drift coefficient and $\zeta(t)$ is a random force due to random density fluctuations of the environment around the particle. For the sake of simplicity the mass of the thermal particle is set to $m = 1$. The random force $\zeta(t)$ can be characterized by its first and second moments as follows.
\begin{align}
  \ave{\zeta(t)} = 0, \quad \ave{\zeta(t)\zeta(t')} = 2 \alpha k_{\rm B} T \delta(t-t').
\end{align}
It is well known that
when $U(v)$ is a parabolic potential $U_s(v) = v^2 /2$, which is equivalent to
Rayleigh dissipation function \cite{Rayleigh}, the drift force in $v$-space obeys the linear
constitutive relation $F_d(v) = - \alpha \frac{d}{dv} U_s(v) = - \alpha v$. In this case a Gaussian pdf
 \begin{align}
  p(v) = \frac{1}{Z(\beta)} \,
     \exp \left[ -\beta \frac{v^2}{2} \right],
  \end{align}
is the stationary state of a thermal particle in this parabolic potential.

\begin{figure}[h]
  \includegraphics[width=.85\linewidth]{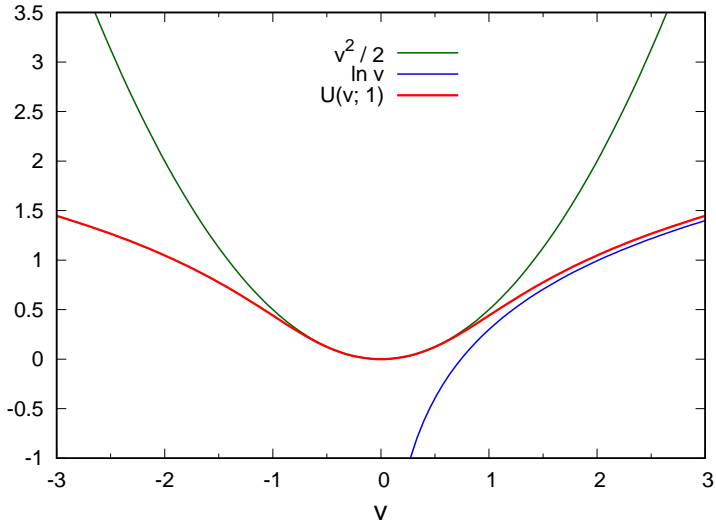}
%\vspace{4mm}
  \caption{ The weakly confining potential $U(v; v_c)$ \eqref{weakU} with the controlling parameter $v_c=1$ (red curve). For the sake of comparison, a parabolic (green curve) and logarithmic (blue curve) potentials are also shown. }
\label{fig1}

\end{figure}

In contrast, we consider the weakly confining potential \eqref{weakU} with a controlling parameter $v_c$.
Figure \ref{fig1} shows this weakly confining potential $U(v; v_c)$ with $v_c = 1$.
For a small $v$ much less than $v_c$, the potential $U(v; v_c)$ is well approximated with $v^2/2$, and
for a large $v$ much lager than $v_c$, the potential $U(v; v_c)$ behaves as $v_c^2 \ln v$, i.e.,
\begin{align}
 U(v; v_c)
     \sim \left\{
  \begin{array}{ll}
       v^2 / 2,  &  \quad (v \ll v_c), \\
       v_c^2 \, \ln v, & \quad (v \gg v_c).
  \end{array} \right.
\label{behaviorU}
\end{align}
The constitutive relation for this weakly confining potential becomes
\begin{align}
K(v; v_c)  := - \alpha \frac{d}{dv} U(v; v_c)
  =  -\frac{\alpha v}{\sqrt{1+\left(\frac{v}{v_c}\right)^4}},
\label{K}
\end{align}
where $K(v; v_c)$ is the nonlinear drift force associated with the potential $U(v; v_c)$.
From the behavior of $U(v; v_c)$ described in Eq. \eqref{behaviorU} we see that for a small $v \ll v_c$ the drift force $K(v; v_c)$ is
well approximated with the linear constitutive relation $K(v; v_c) = - \alpha v$, while
for a larger $v \gg v_c$ $K(v; v_c) \propto -1/v$, which means that the particle cannot be captured in the region $R_c$ around the minimum point of this potential $U(v; v_c)$ if the particle velocity $v$ is much larger than $v_c$.

%%%%%%%%%%%%
Now, we consider the canonical pdf $p(v)$ of a thermal particle 
in the weakly confining potential $U(v; v_c)$,
\begin{align}
  p(v) \propto \exp \left[ -\beta \, U(v; v_c) \right]
  = \exp \left[ -\beta \, \frac{v_c^2}{2} \arsinh \left( \frac{v^2}{v_c^2} \right) \right],
\end{align}
where $\beta$ is the coldness, or inverse temperature $1/(k_{\rm B} T)$.
At this stage let us re-parametrize the controlling parameter $v_c$ as
\begin{align}
   \frac{v_c^2}{2} = \frac{1}{\kappa \beta},
  \label{repara}
\end{align}
where $\kappa$ is a new real parameter. With this reparametrization, the parameter $v_c$ is
a function of $\kappa$ and $\beta$, i.e., $v_c = v_c(\kappa, \beta)$.
We then find that
  \begin{align}
    p(v) &\propto \exp \left[ \frac{1}{\kappa} \arsinh \left( -\kappa \beta \, \frac{v^2}{2} \right) \right]
= \kexp \left[ -\beta \, \frac{v^2}{2} \right],
    \end{align}
where we used the relation \eqref{kexp} in the last step. We thus see that the canonical pdf of a thermal particle in the weakly confining potential is the $\kappa$-deformed Gaussian.

Next we consider the following linear FP equation \cite{W10}
\begin{align}
  \frac{\partial}{\partial t} p(v, t) =
  -\frac{\partial}{\partial v} \Big( K(v; v_c) \, p(v, t) \Big)  
  + D \frac{\partial^2}{\partial v^2} p(v,t),
\label{linFP}
\end{align}
where  $K(v; v_c)$ denotes the nonlinear drift coefficient in Eq. \eqref{K},
and $D$ a constant diffusion coefficient.
The stationary condition $\frac{\partial}{\partial t} p_s(v) =0$ leads to
\begin{align}
  \frac{\partial}{\partial v} \ln p_s(v) &=
 \frac{K(v; v_c)}{D} = -\frac{\frac{\alpha}{D} v}{\sqrt{1+\left(\frac{v}{v_c}\right)^4}} 
%\notag \\
  = -\frac{\alpha v_c^2}{2 D} \, \frac{\partial}{\partial v} \arsinh\left(\frac{v^2}{v_c^2}\right).
\end{align}
This relation becomes
\begin{align}
  \ln p_s(v) &= \frac{\alpha v_c^2}{2 D} \, \arsinh\left(-\frac{v^2}{v_c^2}\right) + \textrm{const.}
\end{align}
We thus obtain the stationary pdf of the FP equation \eqref{linFP} as
\begin{align}
p_s(v) &\propto \exp\left[\frac{\alpha v_c^2}{2 D} 
 \arsinh \left(-\frac{v^2}{v_c^2}\ \right)   \right].
\label{ps1}
\end{align}
Next  we introduce the parameters as
\begin{align}
 \kappa = \frac{2 D}{\alpha v_c^2},\quad
 \beta = \frac{\alpha}{D},
\label{kappa_beta}
\end{align}
which are consistent with the relation \eqref{repara}.
Rewriting Eq. \eqref{ps1} by using these parameters, we have 
\begin{align}
p_s(v) &= \frac{1}{Z(\beta)} \exp\left[\frac{1}{\kappa} 
 \arsinh \left(-\kappa \, \beta \, \frac{ v^2}{2} \right)   \right]
%\notag \\
 = \frac{1}{Z(\beta)} \kexp \left(-\beta \, \frac{v^2}{2} \right),
\label{kexp-rel}
\end{align}
which is the $\kappa$-deformed Gaussian.
In this analysis it is remarkable that: i) as shown in  Eq. \eqref{kappa_beta}, the deformed parameter $\kappa$ is expressed in terms
of the microscopic physical parameters $\alpha, D$ and the controlling parameter $v_c$ which
determines the nonlinear constitutive relation \eqref{K}.
In the limit of $v_c \to \infty$ the deformed parameter $\kappa$ reduces to $0$;
ii) the parameter $\beta$ is expressed as the ratio of
the friction coefficient $\alpha$ to the diffusion coefficient $D$,
in analogy with the fluctuation-dissipation relation;
iii) for the region $\kappa > 2$ of the deformed parameter $\kappa$, the integral $\int dv  p_s(v)$ diverges, and consequently
$p_s(v)$ is no longer a pdf since it cannot be normalized.  By using the above parameters \eqref{kappa_beta}, this condition is equivalently expressed as
$ v_c ^2  < k_B T = 1/\beta = D / \alpha$, which physically means the captured energy $v_c^2 /2$ of the weakly confining potential  $U(v; v_c)$ at $v=v_c$
is smaller than the mean thermal energy $k_B T /2$. Then a thermal particle which has an enough high thermal energy cannot be captured in the region $R_c$ around the minimum of $U(v; v_c)$. In other words, for a fixed value of $v_c$, only the limited value $\beta > 1/v_c^2$ of $\beta$ is allowed for convergence of the pdf $p_s(v)$; 
iv) in the deformed parameter region $2/3 < \kappa$, the second moment
\begin{align}
  \ave{ v^2 } = \int dv \, v^2 p_s(v), 
\end{align}
of the $\kappa$-Gaussian becomes infinite, in which the mean kinetic energy $\ave{v^2} / 2$ diverges, and it is a hall-mark of anomalous transport. We will discuss this point later.
 
By introducing  the Lyapunov functional ${\mathcal F[p]}$ as
\begin{align}
  {\mathcal F[p]} &\equiv U[p] - \frac{D}{\alpha} \, S^{\rm BG}[p],
  \label{F}
\end{align}
with
\begin{align}
  S^{\rm BG}[p] &= -\int dv \; p(v, t) \ln p(v, t), \\
  U[p] &\equiv \int dv \; U(v; v_c) \, p(v, t),
\end{align}
we show \cite{W10} that
\begin{align}
   \frac{d {\mathcal F}}{d t} &= \int dv \; 
       \frac{\partial }{\partial p} \left[ U(v; v_c)\, p
         + \frac{D}{\alpha} \, p \ln p \right] \;  
  \frac{\partial p}{\partial t}  \notag \\
    &= \int dv \; \left[ U(v; v_c)
         + \frac{D}{\alpha} \, ( \ln p + 1 ) 
     \right]
%\notag \\
% & \qquad \qquad \quad \times \,
\frac{\partial}{\partial v}\,\left[-K(v; v_c) p +
   D \frac{\partial p}{\partial v} \right]
\nonumber \\
 & =  -\int dv \, \frac{p}{\alpha}
     \left[ -K(v; v_c) + D \, \frac{\partial}{\partial p} \ln p \right]^2
 \le 0,
\end{align}
which states the time evolution of ${\mathcal F}[p]$ is non-increasing, i.e.,
${\mathcal F}[p]$ is the Lyapunov functional w.r.t. the FP equation \eqref{linFP}.
Consequently the ${\mathcal F}[p]$ is minimized by the stationary solution, which is a $\kappa$-generalized Gaussian.

Having described the FP analysis of a thermal particle in the weakly confining potential $U(v; v_c)$,
we next point out the violation of the standard formula
\begin{align}
  \ave{U(v)} = - \frac{\partial}{\partial \beta} \ln Z(\beta),
  \label{formula}
\end{align}
on the average energy for a canonical pdf.
It is well known that the average energy $\ave{U(v)}$ for a canonical pdf
\begin{align}
 p(v) =  \frac{1}{Z(\beta)} \, \exp \left[-\beta U(v) \right],
\end{align}
is obtained by the famous relation
\eqref{formula}.
In the derivation of this formula we assume that a potential $U(v)$ is independent 
of $\beta$.
However this assumption is no longer valid for the weakly confining potential $U(v; v_c)$,
since $v_c = v_c(\kappa, \beta)$ is a function of $\kappa$ and $\beta$.
This weakly confining potential $U(v; v_c)$ depends not only on $v$ but also on
$\beta$; in other words, $U(v; v_c)$ is \textit{temperature dependent}.
Consequently the above formula \eqref{formula} is no longer valid.
Taking into account the $\kappa$- and $\beta$-dependence of $v_c$, we can obtain
\begin{align}
  \int dv \, \frac{v^2}{2} \frac{p_s(v)}{ \sqrt{ 1 + \left( \frac{v}{v_c} \right)^4}} 
= - \frac{\partial}{\partial \beta} \ln Z(\beta),
 \label{can-rel}
\end{align}
by the straightforward calculations.
The r.h.s. can be easily calculated by using the expression \eqref{kexp-rel} and
the analytical expression \cite{K02} for the $\kappa$-deformed Gaussian integral, as follows
\begin{align}
   Z(\beta) &= \int_{-\infty}^{\infty} dv \, \kexp(-\beta \frac{v^2}{2}) 
\nonumber \\
  &= \sqrt{\frac{\pi}{\beta \kappa}} \,
   \frac{\Gamma(\frac{1}{2 \kappa}-\frac{1}{4})}
  {(1+\frac{\kappa}{2}) \Gamma(\frac{1}{2 \kappa}+\frac{1}{4})} \;
\underset{\kappa \to 0} {\longrightarrow} \; \sqrt{\frac{2 \pi}{\beta}}.
\end{align}
Then 
\begin{align}
   -\frac{\partial}{\partial \beta} \ln Z ( \beta ) = 
  \frac{\partial}{\partial \beta} \ln \sqrt{\beta } =
  \frac{1}{2 \beta}.
\end{align}
The l.h.s. of Eq. \eqref{can-rel} is the half of the following average energy
\begin{align}
   \ave{ v \frac{d}{dv} U(v; v_c)} &=
  \int_{-\infty}^{\infty} dv  \,
  \frac{v^2}{\sqrt{1+\left(\frac{v}{v_c}\right)^4}}
 \, p_s(v).
\label{ave}
\end{align}
By using the relation
\begin{align}
 - \frac{1}{\beta} \frac{d}{dv} \, \kexp\left(-\beta \frac{v^2}{2} \right)
 % = - \frac{1}{\beta} \frac{d}{dv} \, \exp\left(-\beta U_{\kappa \beta}(v^2) \right)
 % = \frac{d U_{\kappa \beta}(v^2)}{dv} \kexp\left(-\beta \frac{v^2}{2} \right).
= \frac{v \kexp\left(-\beta \frac{v^2}{2} \right)}{\sqrt{1+\left(\frac{v}{v_c}\right)^4}},
\end{align}
and integration by part, the r.h.s. of Eq. \eqref{ave} becomes
\begin{align}
-\frac{v}{\beta Z(\beta)}  \kexp \left(-\beta \frac{v^2}{2} \right) \Big\vert_{-\infty}^{\infty}+
\frac{1}{\beta} \int_{-\infty}^{\infty} dv \, p_s(v).
\label{rhs}
\end{align}

Finally, let us consider the case in the parameter region $2/3 < \kappa < 2$, in which the second moment $\ave{v^2}$ of the $\kappa$-deformed Gaussian diverges as stated before.
In this parameter region, the first term in \eqref{rhs} 
becomes zero and the stationary pdf $p_s(v)$ is normalized, we then obtain
\begin{align}
   \ave{ v \frac{d}{d v} U(v; v_c)} = \frac{1}{\beta}=  \frac{D}{\alpha},
\label{this}
\end{align}
which reminds us of \textit{a generalization of equipartition theorem} \cite{To},
\begin{align}
   \ave{ p \frac{\partial}{\partial p} \mathcal{H}} = k_{\rm B} T,
\label{GER}
\end{align}
where $\mathcal{H}$ is the Hamiltonian of a system in thermal equilibrium
with the temperature $T$, and $p$ is a generalized momentum conjugate w.r.t a generalized position $q$. 
Thus, while the mean kinetic energy diverges,
the average energy \eqref{ave} remains finite and is characterized
by the ratio $D/\alpha$ of the diffusion coefficient $D$ to the drift coefficient $\alpha$.
We note that the average energy \eqref{ave} is characterized with the so called un-normalized $\kappa$-escort expectation, or $\kappa$-canonical expectation \cite{MW15}, which is written in our model by
\begin{align}
  {\rm E}_{\kappa, p}[f(v)] &:= \int dv \; f(v)  \frac{d}{dx} \kexp(x) \Big\vert_{x =-\beta \frac{v^2}{2}} 
\notag \\
&=
\int dv \; f(v)  \frac{ \kexp\left(-\beta \frac{v^2}{2} \right)}{\sqrt{1 + \kappa^2 \beta^2 (v^2 / 2)^2}}
 %\notag \\ 
=\int dv \; f(v)  \frac{ \kexp\left(-\beta \frac{v^2}{2} \right)}{\sqrt{1 + \left(\frac{v}{v_c} \right)^4}},
\end{align}
where we used relation \eqref{der_kexp}.
Then, the $\kappa$-canonical expectation of the kinetic energy $v^2/2$ is expressed as
\begin{align}
  {\rm E}_{\kappa, p} \left[ \frac{v^2}{2} \right] 
&= \int dv \; \frac{v^2}{2} \; \frac{ Z(\beta) p_s(v)}{\sqrt{1 + \left(\frac{v}{v_c} \right)^4}} 
= - \frac{\partial}{\partial \beta} Z(\beta),
\label{can_expect}
\end{align}
where the relation \eqref{can-rel} was used in the last step.
In the limit of $\kappa \to 0$, the relation \eqref{can_expect} reduces to the standard relation
\begin{align}
Z(\beta) \ave{ \frac{v^2}{2}} = -\frac{\partial}{\partial \beta} Z(\beta).
\end{align}

%%%%%%%%%%%%%%%%%%%%%%%%%%%%%%%%%%%%%%%%%%%%%%%%%%%%%%%%%%%%%%%%%%%%%5

\section{Conclusion and perspective}

 We have considered a thermal particle in a weakly confining potential $U(v; v_c)$ in $v$-space with a control parameter $v_c$. We have shown that its stationary state is a canonical probability distribution which is the $\kappa$-deformed Gaussian.
In this model the parameter $v_c$ characterizes the $v$-dependency of the weakly confining potential $U(v; v_c)$, and consequently determines the nonlinear constitutive relation \eqref{K}. Then, the deformed parameter $\kappa$ is determined by $v_c$ and $\beta$ as shown
in relations \eqref{kappa_beta}. In this way, we provide a possible physical meaning of the deformed parameter $\kappa$ in this model. Note also that our nonlinear velocity-dependent potential $U(v; v_c)$
can be considered as a generalization of Rayleigh dissipation function with the deformed parameter $v_c$.

Recently, Chaudhuri \cite{Chaudhuri} discussed the stochastic dynamics of the active Brownian particles which are modeled in terms of nonlinear velocity dependent force. Using the associated FP equation, he derived the expression for the total entropy production. Since our model is also described by the nonlinear velocity dependent force, it is worth to further study our model from the view of the entropy production and associated fluctuation theorem.

It is interesting to further study the relation \eqref{can_expect} from the view points of both statistical physics and Information geometry.

Another possible further study is to generalize our model by utilizing Naudts' $\phi$-exponential function \cite{Naudts}.
It is a unification of some deformed-exponential functions. It includes the $q$- and $\kappa$-exponential functions as special cases. 
The inverse function of the $\phi$-exponential function is called $\phi$-logarithmic function, which is defined by
\begin{align}
  \ln_{\phi} (x) \equiv \int_1^x \frac{ds}{\phi(s)}, 
\end{align}
for a positive increasing function $\phi(s)$.
The $\kappa$-logarithm is a special case of Naudt's $\phi$-logarithm
with
\begin{align}
 \phi(s) = \frac{2s}{s^{\kappa}+s^{-\kappa}}.
\end{align}

%\begin{acknowledgements}
%\end{acknowledgements}


%\begin{thebibliography}{}
% etc
%\end{thebibliography}

\end{document}